\documentclass[runningheads]{llncs}
\usepackage{graphicx}
\usepackage{wrapfig}

\usepackage{comment}
\usepackage{url}
\usepackage{amsmath,amssymb,mathrsfs}
\usepackage{mathtools}

\usepackage[utf8]{inputenc}
\usepackage[T1]{fontenc}
\usepackage[english]{babel}

\usepackage{listings}
\usepackage{xspace}
\usepackage{caption}
\usepackage{subcaption}

\usepackage[
 svgnames,
 table,
]{xcolor}

\usepackage[
 colorlinks,
 urlcolor = NavyBlue,
 citecolor = NavyBlue,
 linkcolor = NavyBlue,
]{hyperref}

\newcommand\kw[1]{\lstinline[basicstyle=\sffamily\footnotesize]{#1}}
\def\orcidID#1{\smash{\href{http://orcid.org/#1}{\protect\raisebox{-1.25pt}{\protect\includegraphics{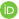}}}}}

\definecolor{codegreen}{rgb}{0,0.6,0}
\definecolor{codegray}{rgb}{0.5,0.5,0.5}
\definecolor{codepurple}{rgb}{0.58,0,0.82}
\definecolor{backcolour}{rgb}{0.95,0.95,0.92}

\lstdefinestyle{mystyle}{
    backgroundcolor=\color{backcolour},   
    keywordstyle=\bf\sffamily,
    numberstyle=\tiny,
    stringstyle=\color{codepurple},
    basicstyle=\sffamily\scriptsize,
    commentstyle=\sl\sffamily,
    breakatwhitespace=false,         
    breaklines=true,                 
    captionpos=b,                    
    keepspaces=true,                 
    numbers=left,                    
    numbersep=5pt,                  
    showspaces=false,                
    showstringspaces=false,
    showtabs=false,                  
    tabsize=2,
    escapechar=@                            
}
\lstdefinelanguage{prism}{
    keywords={root, feature, endfeature, feature, rewards, endrewards, module, modules, endmodule, controller, endcontroller, formula},
    comment=[l]{//}     
}
\begin{document}
\title{Formal Modelling and Analysis \\ of a Self-Adaptive Robotic System}

\author{Juliane P{\"a}{\ss}ler\inst{1}\orcidID{0000-0001-8515-1809} \and
Maurice H. ter Beek\inst{2}\orcidID{0000-0002-2930-6367} \and
Ferruccio Damiani\inst{3}\orcidID{0000-0001-8109-1706} \and\\
S. Lizeth {Tapia~Tarifa}\inst{1}\orcidID{0000-0001-9948-2748} \and
Einar Broch Johnsen\inst{1}\orcidID{0000-0001-5382-3949}}
\authorrunning{J. P{\"a}{\ss}ler et al.}

\institute{University of Oslo, Oslo, Norway\\ \email{\{julipas,sltarifa,einarj\}@ifi.uio.no} \and
ISTI--CNR, Pisa, Italy\\ \email{maurice.terbeek@isti.cnr.it} \and
University of Turin, Turin, Italy\\ \email{ferruccio.damiani@unito.it}}
\maketitle              
\setcounter{footnote}{0} 
\begin{abstract}
  Self-adaptation is a crucial feature of autonomous systems that must cope 
  with uncertainties in, e.g., their environment and their internal state. 
  Self-adaptive systems are often modelled as two-layered systems 
  with a \emph{managed} subsystem handling the domain concerns and a 
  \emph{managing} subsystem implementing the adaptation logic. 
  We consider a case study of a self-adaptive robotic system; more 
  concretely, an autonomous underwater vehicle (AUV) used for pipeline 
  inspection. In this paper, we model and analyse it with the feature-aware probabilistic 
  model checker ProFeat. The functionalities of the AUV are modelled in a 
  feature model, capturing the AUV's variability. This allows us to model 
  the managed subsystem of the AUV as a family of systems, where each 
  family member corresponds to a valid feature configuration of the AUV. 
  The managing subsystem of the AUV is modelled as a control layer capable 
  of dynamically switching between such valid feature configurations, 
  depending both on environmental and internal conditions. We use this model
  to analyse probabilistic reward and safety properties for the AUV.
\end{abstract}

\section{Introduction}
\label{sec:introduction}

Many software systems are subject to different
forms of uncertainty like changes in the surrounding environment,
internal failures and varying user requirements. Often, manually
maintaining and adapting these systems during runtime by a system
operator is prohibitively expensive and error-prone. Enabling systems
to adapt themselves provides several advantages. A system that is able
to perform self-adaptation can also be deployed in environments where,
e.g., communication between an operator and the system is very limited
or impossible, like in space or under water. Thus, self-adaptation
gives a system a higher level of autonomy.
\begin{wrapfigure}[13]{O}{0.42\textwidth}
  \vspace{-6mm}
  \centering
  \includegraphics[width=0.38\textwidth]{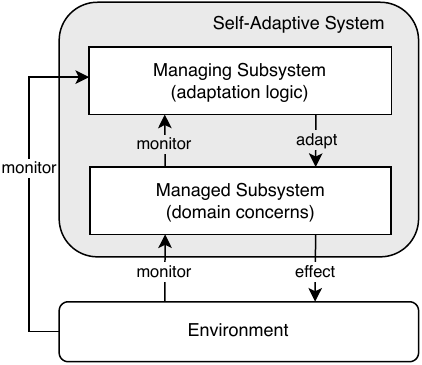}
  \caption{Two-level SAS architecture}
  \label{fig:managed_managing}
\end{wrapfigure}

A self-adaptive system (SAS) can be implemented using a two-layered approach which
decomposes the system into a \emph{managed} and a \emph{managing}
subsystem~\cite{kephartVisionAutonomicComputing2003a}, cf.\ 
Fig.~\ref{fig:managed_managing}.  The \emph{managed} subsystem deals 
with the domain concerns and tries to reach the goals set by the 
system's user, e.g., navigating a robot to a specific location.  The
\emph{managing} subsystem handles the adaptation concerns and defines
an adaptation logic that specifies a strategy on how the system can
fulfil the goals under uncertainty~\cite{weyns2020introduction}, e.g.,
adapting to changing environmental conditions.  While the managed
subsystem may affect the environment via its actions, the managing
subsystem monitors the environment and the internal state of the
managed subsystem. By using the adaptation logic, the managing subsystem
deducts whether and which reconfiguration is needed and adapts the
managed subsystem accordingly. 

This paper models and analyses the case study of a self-adaptive autonomous underwater vehicle (AUV) as a two-layered system. The functionalities of the managed subsystem of the AUV are modelled in a feature model, making the dependencies and requirements between the components of the AUV explicit. The behaviour of the managed subsystem is modelled as a probabilistic transition system whose transitions may be equipped with feature guards, which only allow a transition to be taken if the feature guarding it is included in the current system configuration. Thus, it is modelled as a family of systems whose family members correspond to valid feature configurations. As the behaviour of the AUV depends on environmental and internal conditions, which are both hard to control, we opted for a probabilistic model in which uncontrolled events, like a thruster failure, occur with given probabilities. 
We model the behaviour of the managing subsystem as a control layer that switches between the feature configurations of the managed subsystem according to input from the probabilistic environment model and the managed subsystem.
We consider a simplified version of an AUV, with limited features
and variability, but there are many different possibilities to
extend the model to a more realistic underwater robot.

The case study is modelled in ProFeat~\cite{chrszonProFeatFeatureorientedEngineering2018a}, a tool for probabilistic family-based model checking. 
Family-based model checking provides a means to simultaneously model check, in a single run, properties of a family of models, each representing a different configuration~\cite{TAKSS14}.
Analyses with ProFeat give system operators an estimate of mission duration and the AUV's energy consumption, as well as some safety guarantees.

The main contributions of this paper are as follows:
\begin{itemize}
    \item A case study of an SAS from the underwater robotics domain,
  modelled as a probabilistic feature guarded transition system with dynamic feature switching;
  \item Automated verification of (quantitative) properties that are
  important for roboticists, using family-based analysis.
\end{itemize}

\paragraph{Outline.} 
Sec.~\ref{sec:use_case} presents the case study of pipeline inspection with an AUV. Sec.~\ref{sec:use_case_model} explains both the behaviour of the managed and managing subsystem of the AUV and the environment, as well as their implementation in ProFeat. Sec.~\ref{sec:analysis}
presents quantitative analyses conducted on the case study. Sec.~\ref{sec:related_work} provides related
work. Sec.~\ref{sec:discussion_future_work} discusses our results and ideas for future work.


\section{Case~Study:~Pipeline~Inspection~by~AUV}
\label{sec:use_case}

In this section, we introduce our case study of an AUV used for pipeline inspection, which was inspired by the exemplar
SUAVE~\cite{suave}.

An AUV has the mission to first find and then
inspect a pipeline located on a seabed. During system operation, the
water visibility (i.e., the distance in meters within which the AUV
can perceive objects) might change (e.g., due to currents that swirl
up the seabed), while one or more of the AUV's thrusters might fail
and needs to be restarted before the mission can be continued.

The AUV can choose to operate at three different altitudes,
\emph{low}, \emph{med} (for medium) and \emph{high}. A higher altitude
allows the AUV to have a wider field of view and thus increases its
chances of finding the pipeline during its search. The probability of
a thruster failure is lower at a higher altitude because, e.g.,
seaweed might wrap around the thrusters at a lower altitude.  However,
the altitude at which the AUV can perceive the seabed depends on the
water visibility. With low water visibility, the AUV cannot perceive
the seabed from a high or medium altitude. Thus, it is not always
possible to operate at a high or medium altitude, and the altitude of
the AUV needs to be changed during the search, depending on the
current environmental conditions.
Once the pipeline is found, the AUV will follow it at a low altitude to avoid costs for switching altitudes. In fact, once found, a wider field of view provides no benefit.
However, the AUV can also lose the pipeline again (e.g., when the pipeline was partly covered by sand or the AUV's thrusters failed for some time causing the AUV to drift off its path). In this case, the AUV has to search the pipeline again, enabling all three altitudes.

\paragraph{Two-layered View of the AUV.}
\begin{wrapfigure}[10]{O}{0.5\textwidth}
  \vspace{-10mm}
  \centering
    \includegraphics[width=0.48\textwidth]{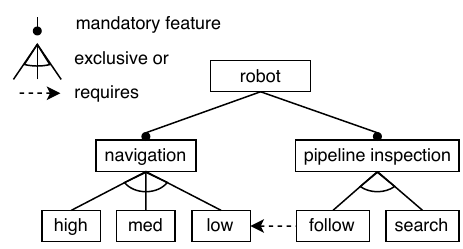}
    \caption{Feature model of the case study}
    \label{fig:feature_diagram}
\end{wrapfigure}
Considering the AUV as a two-layered SAS, the AUV's managed subsystem is responsible for the search for and inspection of the pipeline.
Depending on the current task and altitude of the AUV, a different configuration of the managed subsystem must be chosen. Thus, the managed subsystem can be seen as a family of systems where each family member corresponds to a valid configuration of the AUV. To do so, the different altitudes for navigation (\emph{low}, \emph{med} and \emph{high}) and the tasks \emph{search} and \emph{follow} can be seen as \emph{features} of the managed subsystem that adhere to the feature model in Fig.~\ref{fig:feature_diagram}, which models the dependencies and constraints among the features.
Each configuration of the AUV contains exactly one feature for navigation and one for pipeline inspection, and feature \emph{follow} requires feature \emph{low}, yielding four different configurations of the managed subsystem of the AUV.

The managing subsystem of the case study switches between these configurations during runtime by activating and deactivating the subfeatures of \emph{navigation} and \emph{pipeline inspection}, while the resulting feature configuration has to adhere to the feature model in Fig.~\ref{fig:feature_diagram}. 
The features \emph{low}, \emph{med} and \emph{high} are activated and deactivated according to the current water visibility. If the water visibility is good, 
all three features can be activated; if the water visibility is average, \emph{high} cannot be activated; and if the water visibility is poor, only \emph{low} can be activated.
The managing subsystem switches from the feature \emph{search} to \emph{follow} if the pipeline was found, and from \emph{follow} to \emph{search} if the pipeline was lost.


\section{Modelling the AUV Case Study with  \texorpdfstring{P\MakeLowercase{ro}F\MakeLowercase{eat}}{ProFeat}
 }
\label{sec:use_case_model}

In this section, we describe the behavioural model of the managed and managing subsytem and the environment and model the case study with the family-based
model checker ProFeat\footnote{\url{https://pchrszon.github.io/profeat}.}\,\cite{chrszonProFeatFeatureorientedEngineering2018a}.
ProFeat provides a means to both specify probabilistic system families
and perform family-based quantitative analysis on them. It extends the
probabilistic model checker PRISM\footnote{\url{https://www.prismmodelchecker.org/manual}}\,\cite{kwiatkowskaPRISMVerificationProbabilistic2011}
with functionalities such as family models, features and feature
switches. Thereby, it enables family-based modelling and
(quantitative) analysis of probabilistic systems in which feature
configurations may dynamically change during runtime.
The whole model can be analysed with probabilistic family-based model
checking using PRISM.

Similar to an SAS, a ProFeat model can be seen as a two-layered model,
as illustrated in Fig.~\ref{fig:managed_managing}. The behaviour of
a family of systems that differ in their features, such as the managed
subsystem of an SAS, can be specified. Then a so-called \emph{feature
  controller} can activate and deactivate the features during runtime,
and thus change the behaviour of the system, such as the managing
subsystem of an SAS that changes the configuration of the managed
subsystem. Furthermore, the environment can be specified as a separate module that interacts with the managed and
managing subsystem.  Thus, ProFeat is well suited to model and analyse
the case study described in Sec.~\ref{sec:use_case}.

A ProFeat model consists of three parts: an obligatory feature model
that specifies features and their relations and constraints,
obligatory modules that specify the behaviour of the features, and an
optional feature controller that activates or deactivates features.
The pipeline inspection case study was modelled as a Markov decision
process in ProFeat.\footnote{The ProFeat model is available at \url{https://github.com/JulianePa/auv_profeat}} It consists of (i)~the implementation of the feature model
of Fig.~\ref{fig:feature_diagram}; (ii)~modules describing the
behaviour of the managed subsystem of the AUV (cf.
Fig.~\ref{fig:auv_automaton}) and of the environment (cf. Fig.~\ref{fig:environment}); and (iii)~the
feature controller that switches between features during runtime,
corresponding to the managing subsystem of the AUV (cf.
Fig.~\ref{fig:feature_controller}).  

We start by explaining how the feature model was implemented in
ProFeat in Sec.~\ref{subsec:feature_model_use_case}, then describe the behaviour and implementation of the managed and managing subsystem and of the environment in Sec.~\ref{subsec:managed_use_case}, \ref{subsec:managing_use_case}, and \ref{subsec:environment_use_case} respectively.

\subsection{The Feature Model}
\label{subsec:feature_model_use_case}
We first show how the feature model of the case study is expressed in
ProFeat, including connections and constraints among  features.
Each feature is specified
within a \kw{feature} \ldots\ \kw{endfeature} block, the declaration
of the root feature is done in a \kw{root feature} \ldots\ \kw{endfeature} block.

\paragraph{The Root Feature.}
An excerpt of the implementation of the root feature of the pipeline inspection case study according to Fig.~\ref{fig:feature_diagram} is displayed in Listing~\ref{lst:root_feature}. 
The root feature can be decomposed into subfeatures; in this case only one, the subfeature \kw{robot}, cf.\ Line~\ref{lst:root_feature:robot}. The \kw{all of} keyword indicates that all subfeatures have to be included in the feature configuration if the parent feature, in this case the root feature, is included. It is, e.g., also possible to use the \kw{one of} keyword if exactly one subfeature has to be included.
The modules modelling the behaviour of the root feature are specified after the keyword \kw{modules}. In this case study, the root feature is the only feature specifying modules, thus the behaviour of all features is modelled in the modules \kw{auv} and \kw{environment} described later.

Contrary to an ordinary feature model, ProFeat allows to specify feature-specific
rewards in the declaration of a feature. Like costs, rewards are real values, but unlike costs (and
although they may be interpreted as costs) rewards are meant to
motivate rather than penalise the execution of transitions. Each
reward is encapsulated in a \kw{rewards} \ldots\ \kw{endrewards}
block. In the case study, we consider the rewards \emph{time} and
\emph{energy}, cf.\
Lines~\ref{lst:root_feature:reward_time_beg}--\ref{lst:root_feature:reward_energy_end}
of Listing~\ref{lst:root_feature}. During each transition the AUV
module takes, the reward \kw{time} is increased by~1; it is a
transition-based reward, cf.\
Line~\ref{lst:root_feature:reward_time}. We assume that one time step
corresponds to one minute, allowing us to compute an estimate of a mission's duration.

The reward \kw{energy} is a state-based reward and
can be used to estimate the necessary battery level for a mission completion. If a
thruster of the AUV failed and needs to be recovered, a reward of~2 is
given, 
cf., e.g., Line~\ref{lst:root_feature:rec_high}. The model also reflects that switching between
the search altitudes requires significant energy. Since the altitude
is switched if the AUV is in a search state and a navigation
subfeature that does not correspond to the current search altitude is
active, a higher energy reward is given in these states.
If the AUV needs to switch
between low and high altitude, as, e.g., in
Line~\ref{lst:root_feature:high_to_low}, an energy reward of~4 is
given, while all other altitude switches receive a reward of~2, cf.,
e.g., Line~\ref{lst:root_feature:high_to_med}.  Since the altitude
must be changed to \emph{low} once the pipeline is found, these cases
also receive an energy reward as explained above, cf.\
Lines~\ref{lst:root_feature:found_high}--\ref{lst:root_feature:found_med}.
All other states receive an energy reward of~1.
We use the function \kw{active}
to determine which 
feature is active, i.e., included in the current feature
configuration; given a feature, the function returns true if it is
active and false otherwise.
Note that both \kw{time} and \kw{energy} rewards are interpreted as
costs.

\begin{table}[t]
\begin{lstlisting}[caption=An excerpt of the declaration of the root feature of the case study, label=lst:root_feature]
root feature
	@\label{lst:root_feature:robot}@all of robot;
	modules auv, environment;
	@\label{lst:root_feature:reward_time_beg}@rewards "time"
		@\label{lst:root_feature:reward_time}@[step] true : 1;
	@\label{lst:root_feature:reward_time_end}@endrewards
	@\label{lst:root_feature:reward_energy_beg}@rewards "energy"
		// Costs for being in a recovery state
		@\label{lst:root_feature:rec_high}@(s=recover_high) : 2;
		// .. omitted code ..

		// Costs for switching altitudes 
		@\label{lst:root_feature:high_to_low}@(s=search_high) & active(low) : 4;
		@\label{lst:root_feature:high_to_med}@(s=search_high) & active(med) : 2;
		@\label{lst:root_feature:found_high}@(s=found) & active(high) : 4;
		@\label{lst:root_feature:found_med}@(s=found) & active(med) : 2;
		// .. omitted code ..
	@\label{lst:root_feature:reward_energy_end}@endrewards
endfeature
\end{lstlisting}
\vspace{-2.1\baselineskip}
\end{table}

\paragraph{Ordinary Features.}
The remainder of the feature model is implemented similar to the root feature, but the features do not contain feature-specific modules or rewards. The features are implemented and named according to the feature model in Fig.~\ref{fig:feature_diagram}. To have only one initial state, we initialise the model with the features \kw{search} and \kw{low} active.

\subsection{The Managed Subsystem}
\label{subsec:managed_use_case}

\begin{wrapfigure}[16]{O}{0.6\textwidth}
  \centering
    \includegraphics[width=\linewidth]{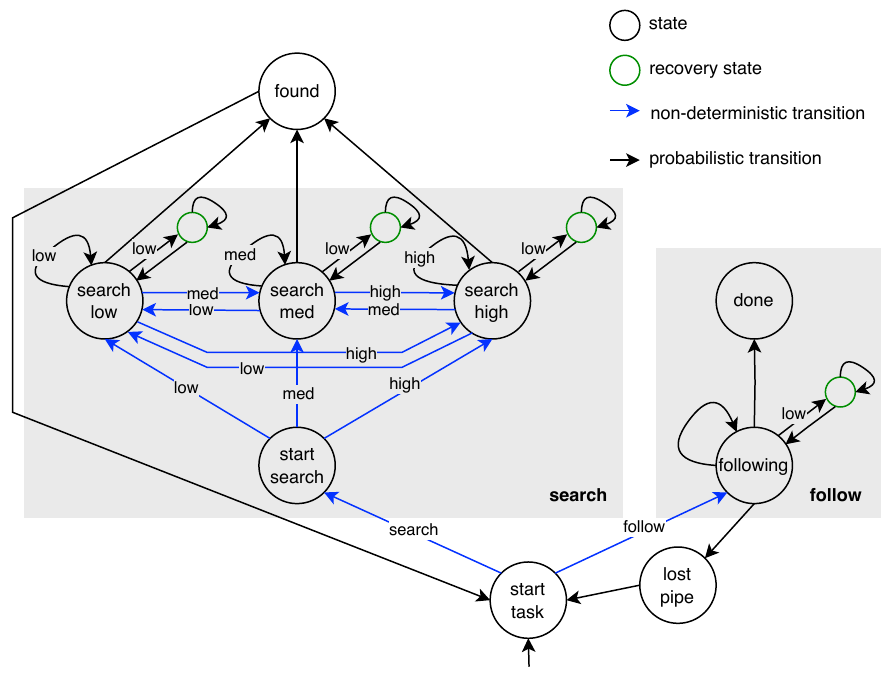}
    \caption{
    The managed subsystem of the AUV
    }
    \label{fig:auv_automaton}
\end{wrapfigure}

\paragraph{The Behavioural Model of the Managed Subsystem.}
The behaviour of the managed subsystem of the AUV can be described by a probabilistic transition system equipped with features that guard transitions (a probabilistic featured transition system). Only if the feature guarding a transition is included in the current configuration of the managed subsystem of the AUV, the transition can be taken. This transition system adheres to the feature model in Fig.~\ref{fig:feature_diagram} and is depicted in 
Fig.~\ref{fig:auv_automaton}, where a number of details have been omitted to avoid cluttering (in particular, all probabilities). The details can be obtained from the publicly available model. The probabilistic model allows to easily model the possibilities of, e.g., finding and losing the pipeline depending on the system configuration.

The transition system can roughly be divided into two parts, one concerning the search for and one the following of the pipeline, as shown by the grey boxes in Fig.~\ref{fig:auv_automaton}. At deployment time, i.e., in state \emph{start task}, the AUV can either immediately start following the pipeline if it was deployed above it, or start searching for it.
During the search for the pipeline, i.e., when the AUV is in the grey
area labelled \emph{search}, the feature \emph{search} should be active and remain active until the state \emph{found} is reached. The managing subsystem can switch between the features
\emph{low}, \emph{med} and \emph{high} during every transition,
depending on the water visibility as described later.
Once the pipeline is found, the managing subsystem has to deactivate the feature \emph{search} and activate the feature \emph{follow}, which also implies activating the feature \emph{low} and deactivating \emph{med} and \emph{high} due to the feature constraints in Fig.~\ref{fig:feature_diagram}. We assume that the managing subsystem activates and deactivates features during transitions, so the features \emph{follow} and \emph{low} should be activated during the transition from the state \emph{found} to the state \emph{start task}. 
When the AUV is following the pipeline, i.e., in the grey area labelled \emph{follow}, it can also lose the pipeline again, e.g., because of sand covering it or because it drifted off its path due to thruster failures. Then the managing subsystem has to activate the feature \emph{search} during the transition from \emph{lost pipe} to \emph{start task}. 

We distinguish two kinds of transitions: probabilistic transitions that model the behaviour of a certain configuration of the managed subsystem (black transitions) and non-deterministic (featured) transitions that depend on the feature choice of the managing subsystem during runtime (blue transitions). The labels \emph{search}, \emph{follow}, \emph{low}, \emph{med} and \emph{high} on the transitions represent the features that have to be active to execute the respective transition.
The non-deterministic (blue) transitions implicitly carry the action to start the task or go to the altitude specified by the feature associated with the transition. For instance, the transitions from \emph{search low} to \emph{search medium} can be taken if the feature \emph{med} is active because the transition has the guard \emph{med}. When taking this transition, the AUV should perform the action of going to a medium altitude.
The probabilistic (black) transitions with a feature label contain the implicit action to stay at the current altitude because the navigation subfeature has not been changed during the previous transition.

Whether a probabilistic or a non-deterministic transition is executed in the search states \emph{search low}, \emph{search medium} and \emph{search high} depends on the managing subsystem, i.e., the controller switching between features (cf.\ Sec.~\ref{subsec:managing_use_case}). If the managing subsystem switched between the features \emph{low}, \emph{med} and \emph{high} during the last transition, a non-deterministic transition to the search state corresponding to the new feature will be executed. Otherwise, a probabilistic transition will be executed. For instance, consider the state \emph{search low}. If the feature \emph{low} is active, then a probabilistic transition will be executed. 
If, however, the managing subsystem deactivated the feature \emph{low} during the last transition and activated either \emph{med} or \emph{high}, then the AUV will perform a transition to the state \emph{search medium} or \emph{search high}, respectively.

\paragraph{The ProFeat Implementation of the Managed Subsystem.}
The module \kw{auv} models the behaviour of the managed subsystem of the AUV as displayed in Fig.~\ref{fig:auv_automaton}, cf.\ Listing~\ref{lst:auv_module} for an excerpt of the model.
As in Fig.~\ref{fig:auv_automaton}, there are thirteen enumerated states in the ProFeat module with names that correspond to the state labels in the figure. The recovery states are named according to the state they are connected to (e.g., the recovery state connected to \kw{search\_high} is called \kw{recover\_high}).
The variable \kw{s} in Line~\ref{lst:auv_module_s} represents the current state of the AUV and is initialised using the keyword \kw{init} with the state \kw{start\_task}.
To record how many meters of the pipeline have already been inspected, the variable \kw{d\_insp} in Line~\ref{lst:auv_module_d_insp} represents the distance the AUV has already inspected the pipeline, it is initialised with~0. The variable \kw{inspect} represents the desired inspection length and can be set by the user during design time.
Since the number of times a thruster failed impacts how much the AUV deviates from its path, the variable \kw{t_failed} can be increased if a thruster fails while the AUV follows the pipeline. It is bounded by the influence a thruster failure can have on the system (\kw{infl_tf}) that can be set by the user during design time.

\begin{table}[t]
\begin{lstlisting}[caption=An excerpt of the ProFeat AUV module of the case study, label=lst:auv_module]
module auv
	@\label{lst:auv_module_s}@s : [0..12] init start_task;
	@\label{lst:auv_module_d_insp}@d_insp : [0..inspect] init 0;
	@\label{lst:auv_module_t_failed}@t_failed : [0..infl_tf] init 0;

	// .. omitted code ..
	// From search state to another state
	@\label{lst:auv_module:search_high_prob_beg}@[step] (s=search_high & active(high)) -> 0.59:(s'=found) 
						@\label{lst:auv_module:search_high_prob_end}@+ 0.4:(s'=search_high) + 0.01:(s'=recover_high);
	@\label{lst:auv_module:search_high_switch_beg}@[step] (s=search_high & active(med)) -> 1:(s'=search_med);
	@\label{lst:auv_module:search_high_switch_end}@[step] (s=search_high & active(low)) -> 1:(s'=search_low);
	// .. omitted code ..
 
	// Following the pipeline
	@\label{lst:auv_module:following_inspect_beg}@[step] (s=following) & (d_insp<inspect) & (t_failed=0)
					-> 0.9: (s'=following) & (d_insp'=d_insp+1)
					@\label{lst:auv_module:following_inspect_end}@+ 0.07: (s'=lost_pipe) + 0.03:(s'=recover_following)
					& (t_failed'=(t_failed<infl_tf? t_failed+1 : t_failed));
	// .. omitted code ..
	@\label{lst:auv_module:following_inspected}@[step] (s=following) & (d_insp=inspect) -> (s'=done);

	// Lost the pipeline
	@\label{lst:auv_module:lost_pipe}@[step] (s=lost_pipe) -> 1: (s'=start_task) & (t_failed'=0);
	// .. omitted code ..
endmodule
\end{lstlisting}
\vspace{-2.1\baselineskip}
\end{table}

The behaviour of the module is specified with \emph{guarded
  commands},  corresponding to possible, probabilistic
transitions, of the form
$$\mbox{\kw{[action] guard -> prob\_1: update\_1 + ... + prob\_n:
    update\_n;}}$$
A command may have an optional label \kw{action} to annotate it or to
synchronise with other modules. In PRISM, the \kw{guard} is a
predicate over global and local variables of the model, which can also
come from other modules. ProFeat extends the guards by, e.g., enabling
the use of the function \kw{active}.  If the guard is true, then the
system state is changed with probability \kw{prob\_i} using
\kw{update\_i} for all $i$. An update describes how the system should
perform a transition by giving new values for variables, either
directly or as a function using other variables.

For instance, consider the command in
Lines~\ref{lst:auv_module:search_high_prob_beg}--\ref{lst:auv_module:search_high_prob_end},
which can be read as follows. If the system is in state
\kw{search\_high} and the feature \kw{high} is active, then with a
probability of~0.59, the system changes its state to \kw{found}, with
a probability of~0.4 it changes to \kw{search\_high} and with a
probability of~0.01 it changes to \kw{recover\_high}. These are
exactly the probabilistic transitions shown in
Fig.~\ref{fig:auv_automaton} exiting from state \emph{search
  high}. This command also has an action label, \kw{step}. Using this
action label, it synchronises with the environment module and the
feature controller, as described later.  The non-deterministic
transitions exiting state \emph{search high} in
Fig.~\ref{fig:auv_automaton} are modelled in
Lines~\ref{lst:auv_module:search_high_switch_beg}--\ref{lst:auv_module:search_high_switch_end}. If
the model is in state \kw{search\_high}, but the feature \kw{low} or
\kw{med} is active, indicating that the AUV should go to the
respective altitude, then the state is changed to the respective
search state.
The transitions exiting the states \kw{search\_med} and \kw{search\_low} are modelled similarly. However, the probability of going to the state \kw{found} is highest from state \kw{search\_high} and lowest from 
\kw{search\_low} because the AUV has a wider field of view when performing the search at a higher altitude. Furthermore, the probability of a thruster failure, i.e., of going to the respective \kw{recover} state, is highest in state \kw{search\_low} and lowest in state \kw{search\_high} because the probability of seaweed getting stuck in the thrusters is higher at a lower altitude.

From the \kw{following} state, the transitions that can be taken depend on the variables \kw{d\_insp} and \kw{t_failed}. 
Lines~\ref{lst:auv_module:following_inspect_beg}--\ref{lst:auv_module:following_inspect_end} consider the case where the distance of the pipeline that has already been inspected \mbox{(\kw{d\_insp})} is less than the distance the pipeline should be inspected (\kw{inspect}) and the variable \kw{t_failed} is 0, indicating that there were no recent thruster failures. 
Then the AUV stays in the \kw{following} state and inspects the pipeline one more meter, it loses the pipeline, or a thruster fails and it transitions to the failure state and increases \kw{t_failed} if \kw{t_failed} is not at its maximum.
If \kw{d\_insp} is less than \kw{inspect} and \kw{t_failed} is greater than~0, the probabilities of following and of losing the pipeline depend on the value of \kw{t_failed}. The bigger the value, the more likely it is to lose the pipeline because it indicates that the AUV's thrusters did not work for some time, causing it to drift off its path.
If the already inspected distance is equal to the required inspection distance, the AUV transitions to the \kw{done} state (cf.\ Line~\ref{lst:auv_module:following_inspected}) and finishes the pipeline inspection.
If the AUV lost the pipeline (cf.\ Line~\ref{lst:auv_module:lost_pipe}), then a transition to \kw{start_task} is taken and the variable \kw{t_failed} is set to~0 again.

All commands in the module \kw{auv} are labelled with \kw{step}. Thus, every transition receives a \kw{time} reward of~1, i.e., the time advances with every transition the AUV takes, cf.\ Lines~\ref{lst:root_feature:reward_time_beg}--\ref{lst:root_feature:reward_time_end} of Listing~\ref{lst:root_feature}.

\subsection{The Environment}
\label{subsec:environment_use_case}
\paragraph{The Behavioural Model of the Environment.}
\begin{wrapfigure}[6]{O}{0.68\textwidth}
  \vspace{-8mm}
    \centering
    \includegraphics[width=0.68\textwidth]{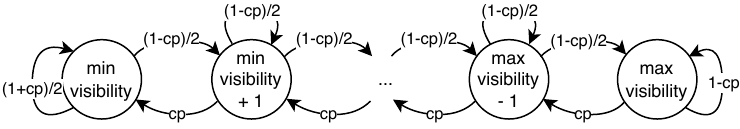}
    \caption{The behaviour of the environment
    }
    \label{fig:environment}
\end{wrapfigure}
We assume that there is a minimum and a maximum visibility of the environment, depending on where the AUV is deployed and set by the user during design time. Furthermore, different environments also have different probabilities of currents that influence the water visibility. This can also be set during design time. 
The behaviour of the environment is then modelled as 
depicted in Fig.~\ref{fig:environment}, where \emph{cp} represents
the \emph{current probability}. With the probability of currents \emph{cp}, the water visibility decreases by~1, while it stays the same or increases by~1 with probability~(1-\emph{cp})/2.
If the water visibility is already at minimum visibility,
the water visibility stays the same with probability
(1+\emph{cp})/2 and, at maximum visibility,
it stays the same with probability (1-\emph{cp}).

\paragraph{The Implementation of the Environment in ProFeat.}
The environment is modelled in a separate \kw{environment} module, cf.\
Listing~\ref{lst:environment_module}.  The variable \kw{water\_visib}
in Line~\ref{lst:environment_module:wv} reflects the current water
visibility and 
is initialised parametrically, depending on the
minimum and maximum visibility, cf.\
Line~\ref{lst:environment_module:init}. The function \kw{round()} is
pre-implemented in the PRISM language and rounds to the nearest
integer.  
The environment module synchronises with the AUV module via the label
of its action, \kw{step}. Since the guard of the only action in the
environment module is \kw{true}, the environment executes a transition
every time the AUV module does. 
By decoupling the environment module from the AUV module, we obtain a
separation of concerns which makes it easier to change the model of
the environment if needed.
\begin{table}[t]
\begin{lstlisting}[caption=The ProFeat environment module of the case study, label=lst:environment_module]
module environment
	@\label{lst:environment_module:wv}@water_visib : [min_visib..max_visib] 
						@\label{lst:environment_module:init}@init round((max_visib-min_visib)/2);
	[step] true -> current_prob: (water_visib'= (water_visib=min_visib?
						min_visib:water_visib-1)) + (1-current_prob)/2: (water_visib'= 
						(water_visib=max_visib? max_visib:water_visib+1))
						+ (1-current_prob)/2: true;
endmodule
\end{lstlisting}
\vspace{-2.1\baselineskip}
\end{table}

\subsection{The Managing Subsystem}
\label{subsec:managing_use_case}

\begin{wrapfigure}[18]{O}{0.68\textwidth}
  \centering
    \includegraphics[width=0.68\textwidth]{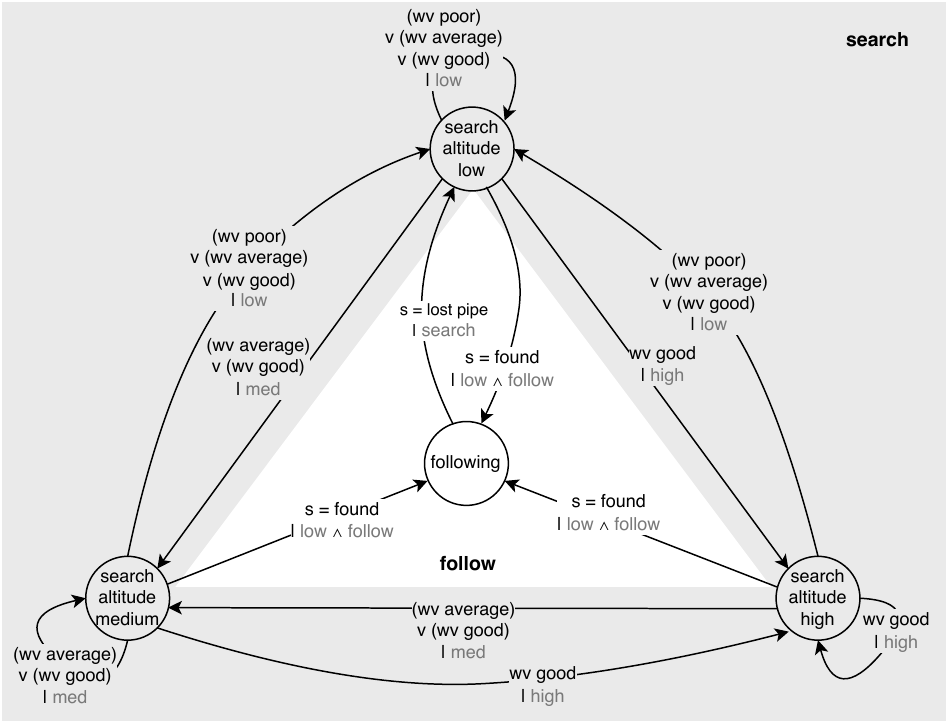}
    \caption{
    The managing subsystem of the AUV
    }
    \label{fig:feature_controller}
\end{wrapfigure}

\paragraph{The Behavioural Model of the Managing Subsystem.}
As described in Sec.~\ref{sec:use_case}, the managing subsystem of the AUV implements the AUV's adaptation logic, which corresponds to activating and deactivating the features of the managed subsystem.
The behaviour of the managing subsystem of the AUV is displayed in
Fig.~\ref{fig:feature_controller}. The grey area of the figure
includes the transitions that can be taken during the search for the
pipeline, and the white area the transitions once the pipeline has been
found.  Each transition contains a guard, written in black, and an
action, written in grey after a vertical bar.

During the search for the pipeline, i.e., in the grey area of Fig.~\ref{fig:feature_controller}, the managing subsystem activates and deactivates the features \emph{low}, \emph{med} and \emph{high} according to the current water visibility as described in Sec.~\ref{sec:use_case}. 
The activated feature is displayed in grey on the transition, implicitly the other two subfeatures of \emph{navigation} are deactivated. Note that the transitions in the grey area implicitly carry the guard \emph{s != found}, i.e., the AUV is not in the state \emph{found}, because they represent the transitions during the search for the pipeline. This guard was omitted for better readability.

Once the pipeline has been found, i.e., the managed subsystem is in the state \emph{found}, one of the transitions in the white area, guarded by $s = \textit{found}$, is taken. These transitions include the action of activating \emph{low} and \emph{follow}, and thus deactivating \emph{med}, \emph{high} and \emph{search}. When the AUV loses the pipeline, i.e., it is in the state \emph{lost pipe}, the managing subsystem activates \emph{search} and deactivates \emph{follow}. Since the AUV is following the pipeline at a low altitude, the AUV will start searching at a low altitude.

\paragraph{The Implementation of the Managing Subsystem in ProFeat.}
The managing subsystem of the AUV is implemented as a feature controller in ProFeat. 
The feature controller can also use \emph{commands} to change the state of the system. Such commands are similar to those used in a module; they are mostly of the form \kw{[action] guard -> update}. Each command can have an optional label \kw{action} to synchronise with the modules, and its \kw{guard} is a predicate of global and local variables of the model and can also contain the function \kw{active}. In contrast to the commands in the modules, the feature controller can \kw{activate} and \kw{deactivate} features in the \kw{update} of a command. Several features can be activated and deactivated at the same time, but this cannot be done probabilistically and the resulting feature configuration has to adhere to the feature model.

In the pipeline inspection case study, subfeatures of \kw{navigation} (i.e., the different altitudes at which the AUV can operate) and subfeatures of \kw{pipeline\_inspection}  (i.e., the tasks the robot has to fulfil) can be switched by the feature controller during runtime, cf.\ Listing~\ref{lst:feature_controller}.

When the feature \kw{search} is active and the pipeline has not been found yet, the feature controller activates and deactivates the altitudes non-deterministically, but according to the current water visibility, as described before. 
The minimum and maximum water visibility can be set by the user during design time and influence the altitudes associated with the features \kw{low}, \kw{med} and \kw{high}; i.e., it influences when the feature controller is able to switch features. To reflect this, the variables \kw{med\_visib} and \kw{high\_visib} are declared as in Lines~\ref{lst:feature_controller:formula1}--\ref{lst:feature_controller:formula2} (a \emph{formula} in PRISM and ProFeat can be used to assign an identifier to an expression).
If the water visibility is less than~\kw{med\_visib}, the feature controller activates \kw{low} (cf.\ Lines~\ref{lst:feature_controller:wv_le_2_beg}--\ref{lst:feature_controller:wv_le_2_end}) because the AUV cannot perceive the seabed from a higher altitude. If the water visibility is between \kw{med\_visib} and \kw{high\_visib}, it chooses non-deterministically between \kw{low} and \kw{med}, whereas it chooses non-deterministically between all three altitudes if the water visibility is above \kw{high\_visib}. Note that it is also possible to deactivate or activate a feature if it is already inactive or active, respectively.

When the pipeline is found, i.e., the AUV is in state \kw{found}, the feature controller activates the feature \kw{follow} and deactivates \kw{search}, cf.\ Lines~\ref{lst:feature_controller:found_beg}--\ref{lst:feature_controller:found_end}. Since the AUV should be at a low altitude while following the pipeline, the feature controller also deactivates the features \kw{high} and \kw{med} and activates \kw{low}. If the AUV lost the pipeline, i.e., it is in state \kw{lost\_pipe}, the feature controller deactivates \kw{follow} and activates \kw{search} to start the search for the pipeline, cf.\ Lines~\ref{lst:feature_controller:lost_pipe_beg}--\ref{lst:feature_controller:lost_pipe_end}.

The feature controller synchronises with the \kw{auv} and \kw{environment} modules via action label \kw{step}. Since all transitions of the modules and feature controller have the same action label, they can only execute a transition if there is a transition with a guard evaluating to true in both modules and in the feature controller. Thus, the feature controller needs to include a transition doing nothing if the feature \kw{follow} is active and the AUV is not in state \kw{lost\_pipe}, cf.\ Line~\ref{lst:feature_controller:follow}.

\begin{table}[t]
\begin{lstlisting}[caption=An excerpt of the ProFeat feature controller of the case study, label=lst:feature_controller]
@\label{lst:feature_controller:formula1}@formula med_visib = (max_visib-min_visib)/3;
@\label{lst:feature_controller:formula2}@formula high_visib = 2*(max_visib-min_visib)/3;

controller
	// Change altitude depending on water visibility
	@\label{lst:feature_controller:wv_le_2_beg}@[step] (s!=found) & active(search) & water_visib<med_visib
						@\label{lst:feature_controller:wv_le_2_end}@-> activate(low) & deactivate(high) & deactivate(med);
	// .. omitted code ..
      
	// Switch task from "search" to "follow"
	@\label{lst:feature_controller:found_beg}@[step] (s=found) & active(search) -> deactivate(search)&activate(follow) 
						@\label{lst:feature_controller:found_end}@& activate(low) & deactivate(med) & deactivate(high);
      
	// Switch task from "follow" to "search"
	@\label{lst:feature_controller:lost_pipe_beg}@[step] (s=lost_pipe) & active(follow)
						@\label{lst:feature_controller:lost_pipe_end}@-> deactivate(follow) & activate(search);
      
	// Enable transitions when following the pipeline
	@\label{lst:feature_controller:follow}@[step] (s!=lost_pipe) & active(follow) -> true;
endcontroller
\end{lstlisting}
\vspace{-2.1\baselineskip}
\end{table}


\section{Analysis}
\label{sec:analysis}

ProFeat automatically converts models to PRISM for probabilistic model checking. 
To analyse a PRISM model,
properties can be specified in the PRISM property specification language, which includes several probabilistic temporal logics like PCTL, CSL and probabilistic LTL. For family-based analysis, ProFeat extends this specification language to include, e.g.,
the function \kw{active}. (ProFeat constructs have to be specified in
\kw{\$\{...\}} to be correctly translated to the PRISM property
specification language.)

The operators used for analysis in this paper are \kw{P} and \kw{R}, which reason about probabilities of events
and about expected rewards, respectively.  Since we use Markov
decision processes which involve non-determinism, these operators must be further specified to ask for the \emph{minimum} or
\emph{maximum} probability and expected cost, respectively, for all
possible resolutions of non-determinism.

The analysis of the model considered two different aspects. First, the rewards \kw{energy} and \kw{time} were used to compute some safety guarantees that can be used for the deployment of the AUV. Second, safety properties with regard to unsafe states were analysed. Note that it is not necessary to analyse whether the model satisfies the constraints of the feature model because this is automatically ensured by ProFeat.
\begin{wraptable}[4]{O}{0.7\textwidth}
\vspace{-2.5\baselineskip}
  \centering
  \footnotesize
    \caption{Two different scenarios used for analysis}\vspace*{-.5\baselineskip}
    \resizebox{0.68\textwidth}{!}{
    \begin{tabular}{|c||c|c|c|c|}
        \hline
        \textbf{Scenario}   & \textbf{min\_visib}   & \textbf{max\_visib}   & \textbf{current\_prob}    & \textbf{inspect} \\
        \hline
        1 (North Sea)                  & 1                     & 10                    & 0.6                       & 10\\
        \hline
        2 (Caribbean Sea)                  & 3                     & 20                    & 0.3                       & 30\\
        \hline
    \end{tabular}
    }
    \label{tab:scenarios}
\end{wraptable}

We analysed two different scenarios; the values used in these scenarios are reported in Table~\ref{tab:scenarios}. Scenario~1 is in the North Sea, where the minimum and maximum water visibility (in 0.5~meter units)  are relatively low and the probability of currents that decrease the water visibility is relatively high. In this case, only 10~meters of the pipeline have to be inspected.
Scenario~2 is in the Caribbean Sea, with a higher minimum and maximum visibility and a lower probability of currents compared to the North Sea, and 30~meters of pipeline that have to be inspected.
For both scenarios, we first analysed whether it is always possible to finish the pipeline inspection, i.e., reach the state \kw{done}. This could be confirmed since the minimum probability for all resolutions of non-determinism of eventually reaching the state \kw{done} is~1.0.

\paragraph{Reward Properties.}
\begin{wraptable}[8]{O}{0.4\textwidth}
  \vspace{-8mm}
    \centering
    \caption{Expected min-/max\-imum rewards for completing the mission for both scenarios}
    \resizebox{0.38\textwidth}{!}{
    \begin{tabular}{|c|c|c|c|c|}
        \hline
        \textbf{}         & \multicolumn{2}{c|}{\textbf{Energy}} & \multicolumn{2}{c|}{\textbf{Time}} \\\hline
        \textbf{Scenario} & \textbf{min}     & \textbf{max}     & \textbf{min}    & \textbf{max}    \\\hline
        1                 & 24.78            & 44.39            & 23.66           & 32.40           \\\hline
        2                 & 59.08            & 4723.29         & 55.54           & 1315.58          \\\hline
    \end{tabular}%
    }
    \label{tab:rewards}
\end{wraptable}
The rewards \kw{time} and \kw{energy} were used to analyse some safety properties related to the execution of the AUV. Since the AUV only has a limited amount of battery, an estimation of the energy needed to complete the mission is required. This ensures that the AUV is only deployed for the mission if it has sufficient battery to complete it. The commands in Listing~\ref{lst:analysis_rewards} were used to compute the minimum and maximum expected energy (for all resolutions of non-determinism) to complete the mission. Since the model includes two reward structures, the name of the reward has to be specified in \kw{\{"..."\}} after the \kw{R} operator.
Similarly, the minimum and maximum expected time to complete the mission was analysed to give the system operators an estimate of how long the mission might take. The results for Scenarios~1 and~2 are reported in Table~\ref{tab:rewards}. It can be seen that the variation of the parameters in the two scenarios strongly influences the expected energy and time of the mission.
It is interesting to see the difference between minimum and maximum expected energy and minimum and maximum expected time for Scenario~2 are significantly bigger than for Scenario~1. In particular, the maximum expected energy and time are much higher for Scenario~2 than for Scenario~1. Further analysis in this direction could investigate trade-offs between different scenarios and a better understanding of the influence in the results for the different parameters. 

\begin{table}[t]
\begin{lstlisting}[caption=Analysis using the rewards, label=lst:analysis_rewards]
R{"energy"}min=? [F ${s=done}];
R{"energy"}max=? [F ${s=done}];
\end{lstlisting}
\vspace{-2.1\baselineskip}
\end{table}

\begin{table}[t]
\begin{lstlisting}[caption=Analysis of unsafe states, label=lst:analysis_unsafe]
@\label{lst:analysis_unsafe:unsafe}@label "unsafe" = s=recover_high | s=recover_med | s=recover_low
						| s=recover_following;
label "safe" = s=start_task | s=lost_pipe | s=start_search | s=search_high 
						@\label{lst:analysis_unsafe:safe}@| s=search_med | s=search_low | s=found | s=following |s=done;
@\label{lst:analysis_unsafe:all_safe}@Pmin=? [G "safe"]; 
@\label{lst:analysis_unsafe:soon_safe_min}@filter(min, Pmin=? [ F<=k "safe" ], "unsafe");
@\label{lst:analysis_unsafe:soon_unsafe_max}@filter(max, Pmax=? [ F<=k "unsafe" ], "safe");
@\label{lst:analysis_unsafe:soon_unsafe_avg}@filter(avg, Pmax=? [ F<=k "unsafe" ], "safe");
\end{lstlisting}
\vspace{-2.1\baselineskip}
\end{table}

\paragraph{Unsafe States.}
Thruster failures, although we assume that they can be repaired, pose a threat to the AUV. Unforeseen events like strong currents might cause the AUV to be damaged, e.g., by causing it to crash into a rock. To analyse this, the state space was partitioned into two parts, \emph{safe} and \emph{unsafe} states. This was achieved by using labels, cf.\ Lines~\ref{lst:analysis_unsafe:unsafe}--\ref{lst:analysis_unsafe:safe} of Listing~\ref{lst:analysis_unsafe}.

These labels were then used to calculate the probability of several properties. The minimum probability of only taking safe states (cf.\ Line~\ref{lst:analysis_unsafe:all_safe}) was shown to be~0.65 for Scenario~1 and~0.32 for Scenario~2. As expected, the probability of only taking safe states is higher for a shorter pipeline inspection.
It is also important to ensure that a safe state will be reached from an unsafe state after a short period of time, as, e.g., in Line~\ref{lst:analysis_unsafe:soon_safe_min}, where \kw{k} is an integer. For every unsafe state, the minimum probability (for all possible resolutions of non-determinism) of reaching a safe state within \kw{k}~time steps is calculated. Then the minimum over all these probabilities is taken. Thus, it gives the minimum probability of reaching a safe state from an unsafe state in \kw{k}~time steps. PRISM experiments allow analysing this property automatically for a specified range of \kw{k}.
Using PRISM experiments, it was shown that in both scenarios the probability of reaching a safe state from an unsafe state is above~0.95 after 5~time steps and above~0.99 after 7~time steps.

\begin{figure}
  \centering
    \includegraphics[width=0.68\textwidth]{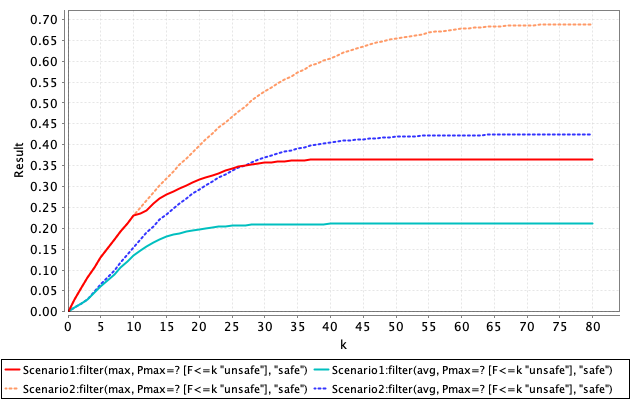}
    \caption{Results 
    for reaching an unsafe state from a safe state in \kw{k} time steps}
    \label{fig:safe_unsafe}
  \end{figure}

  The probability of going to an unsafe state from a safe state should be as small as possible. This is analysed with the properties in Lines~\ref{lst:analysis_unsafe:soon_unsafe_max}--\ref{lst:analysis_unsafe:soon_unsafe_avg}. First, the maximum probability (over all possible resolutions of non-determinism) for reaching an unsafe state from a safe state is calculated, and then the maximum (or average) is taken. Again, PRISM experiments were used to analyse this, the plotted graphs for Scenarios~1 and 2 are displayed in Fig.~\ref{fig:safe_unsafe}. 
  They show that the probability of reaching an unsafe state from a safe state increases with the number of considered time steps. Furthermore, the probability of reaching an unsafe state from a safe state stabilises much later and at a higher value in Scenario~2 than in Scenario~1. While the maximum probability of reaching an unsafe state from a safe state stabilises after about 42~time steps at $\approx$0.37 in Scenario~1, it stabilises after about 76~time steps at $\approx$0.69 in Scenario~2. Similar differences can be observed for the average probability.

\section{Related Work}
\label{sec:related_work}

The analysis of behavioural requirements is often crucial when
developing an SAS that operates in the uncertainty of a physical
environment. These requirements often use quantitative metrics that
change during runtime. 
Both rule-based and goal-based
adaptation logics can be used to enable the SAS to meet its behavioural
requirements.  Many practitioners rely on formal methods to provide
evidence for the system's compliance with such
requirements~\cite{weynsSurvey2012,luckcuckSurvey2019}, but many
different methods are
used~\cite{hezavehiUncertaintySelfadaptiveSystems2021,AMV23}.
We consider related work for family-based modelling and analysis approaches.

Family-based model checking of transition systems with features
allows to model check properties of multiple behavioural models in a single run, following the seminal work by Classen et al.~\cite{CHSLR10}.
Such model-checking tools can be encoded in well-known classical model checkers like SPIN~\cite{Hol04}, NuSMV~\cite{CCGGPRST02} or PRISM~\cite{kwiatkowskaPRISMVerificationProbabilistic2011}. 
In this paper, we used ProFeat~\cite{chrszonProFeatFeatureorientedEngineering2018a}, a software tool built on top of PRISM for the analysis of feature-aware probabilistic models. Alternatively, QFLan~\cite{VBLL18} offers probabilistic simulations to yield statistical approximations, thus trading 100\% precision for scalability.
In~\cite{chrszon20splc,chrszon23jss}, configurable systems are modelled and analysed as role-based systems, an extension of feature-oriented systems, with a focus on feature interaction; in contrast to our paper, they do not consider a separation between managed and managing subsystem. 

Software product lines (SPLs) can be seen as 
families of (software product) models where feature selection yields variations in the products (configurations).
SPLs have previously been proposed to model
static variability, i.e., variability during design time, for robotic
systems~\cite{gherardiModelingReusingRobotic2014}.  
In~\cite{brugaliSoftwareProductLine2021}  it is argued that most of the costs for robotic systems come from non-reusable software. A robotic system mostly contains software tailored to the specific application and embodiment of the robot, and often even software libraries for common robotic functionalities are not reusable. Therefore, they must be re-developed all the time. 
Thus, a new approach for the development of robotic software using SPLs is proposed in~\cite{brugaliSoftwareProductLine2021}. 

Finally,
dynamic SPLs (DSPLs)~\cite{hallsteinsenDynamicSoftwareProduct2013,hincheyBuildingDynamicSoftware2012} have been proposed to manage
variability during runtime for self-adaptive
robots~\cite{brugaliDynamicVariabilityMeets2015}.  
There are several approaches that model, but do not analyse, SASs as DSPLs, e.g., \cite{bencomo08splcw,dhunganaDomainspecificAdaptationsProduct2007,hallsteinsenUsingProductLine2006}.
For robotics, the authors in~\cite{gherardiModelingReusingRobotic2014} propose the toolchain HyperFlex to model robotic systems as SPLs; it supports the design and reuse of reference architectures for robotic systems and was extended with the Robot Perception Specification Language for robotic perception systems in~\cite{brugaliManagingFunctionalVariability2017}. It allows to represent variability at different abstraction levels, and feature models from different parts of the system can be composed in several different ways. However, contrary to the approach used in this paper, Hyper\-Flex only considers design time variability. Furthermore, it is only used for modelling robotic systems, not for analysing them.


\vspace*{-.25\baselineskip}
\section{Discussion and Future Work}
\label{sec:discussion_future_work}
In this paper, we used a feature model together with a probabilistic, feature guarded transition system to model the managed subystem of an AUV used for pipeline inspection, and a controller switching between these features to model the managing subsystem of the AUV.
This allowed modelling the managed subsystem of the AUV  as a family of systems, where each family member corresponds to a valid feature configuration of the AUV. The managing subsystem could then be considered as a control layer capable of dynamically switching between these feature configurations depending on both environmental and internal conditions.
The tool ProFeat was used for probabilistic family-based model checking, analysing reward and safety properties.

ProFeat allowed to model the two different layers of abstraction of an SAS, the managed and managing subsystem, which also makes it easier to understand the model and the adaptation logic. Furthermore, it makes analysing all configurations of the managed subsystem more efficient by enabling family-based model checking. However, it remains to be seen how this scales with larger models.

The case study in this paper is of course a highly simplified model of
an AUV and its mission. However, we showed that it is feasible to model and analyse a two-layered self-adaptive cyber-physical system as a family of configurations with a controller switching between them.
To analyse a real AUV, both the models of the AUV and the environment, and in particular the probabilities, have to be adapted to the robot and the environment with the help of real data and domain experts. We plan to investigate this together with an industrial partner of the MSCA network REMARO (Reliable AI for Marine Robotics).

In the future, we plan to investigate which kind of models can be modelled and analysed as we did with the case study to try to find a general methodology for modelling and analysing SASs as family-based systems.
Furthermore, we plan to find optimal strategies for the managing
subsystem, i.e., the controller switching between features, e.g., to
minimise energy consumption. We would also  like to find patterns
between choosing a certain feature configuration and the effect of
this on quality criteria of the system. Finding such control  patterns
could help to improve the adaptation logic of the managing subsystem
to be more resilient towards faults.


\paragraph{\rm\bf Acknowledgments.}
\begin{small}
We would like to thank Clemens Dubslaff for explaining ProFeat and its usage to us, and for answering numerous questions. Furthermore, we would like to thank Rudolf Schlatte for his help in preparing the artifact for the final artifact submission.
This work was supported by the European Union’s Horizon 2020 Framework Programme through the MSCA network REMARO (Grant Agreement No 956200), by the Italian project NODES (which has received funding from the MUR – M4C2 1.5 of PNRR with grant agreement no. ECS00000036) and by the Italian MUR PRIN 2020TL3X8X project T-LADIES (Typeful Language Adaptation for Dynamic, Interacting and Evolving Systems).
This preprint has not undergone peer review or any post-submission improvements or corrections. The Version of Record of this contribution is published in Lecture Notes in Computer Science, vol 14300, and is available online at \url{https://doi.org/10.1007/978-3-031-47705-8_18}.
\end{small}

\bibliographystyle{splncs04}
\bibliography{references}

\end{document}